\documentclass[11pt]{article}
\usepackage{iap2000,epsfig,rotate}

\def\be{\begin{equation}}
\def\ee{\end{equation}}
\def\bea{\begin{eqnarray}}
\def\eea{\end{eqnarray}}

\def\hmpc{~h$^{-1}$ Mpc~}

\begin{document}
\vspace*{4cm}
\title{ The VIOLENT ENVIRONMENT of the SHAPLEY CONCENTRATION: \\
        a MULTIWAVELENGTH VIEW }

\author{ SANDRO BARDELLI \& ELENA ZUCCA \\
         (bardelli@bo.astro.it, zucca@bo.astro.it) }

\address{ Osservatorio Astronomico di Bologna, via Ranzani 1, 
          I-40127 Bologna, Italy \\
{\rm In collaboration with: \\
 T.~VENTURI, R.~MORGANTI, R.W.~HUNSTEAD (radio) \\
 S.~ETTORI, S.~DE GRANDI, S.~MOLENDI, G.~ZAMORANI (X-rays) \\
 A.~BALDI (optical) 
}
}

\maketitle\abstracts{
Rich superclusters are the ideal environment for the detection of cluster
mergings, because the high peculiar velocities induced by the enhanced
local density of the large-scale structure favour the cluster-cluster
collisions, in the same way as seen in the simulations.
\\
The Shapley Concentration supercluster represents a unique laboratory where
it is possible to follow cluster mergings and to test related astrophysical
consequences, as the formation of shocks, radio halos, relics and 
wide angle tail radiosources, and the presence of galaxies with enhanced
star formation. 
\\
We present the results of an extensive multiwavelength survey of the central
part of the Shapley Concentration, with the use of optical spectra,
radio and X-ray data. 
}

%
\section{Introduction}

Merging clusters are the most energetic and common phenomena in the Universe,
but it is not clear in which way the kinetic energy of this phenomenon
is dissipated and which is the influence on the galaxy population.   
There are some observational features that have been associated with
the cluster merging, like shocks in the hot gas, radio halos, relics and 
wide angle tail radiosources, and the presence of starburst galaxies, even if
the precise theorical description is not yet completely assesed. 
\\
Rich superclusters are the ideal environment for the detection of cluster 
mergings, because the peculiar velocities induced by the enhanced local 
density of the large-scale structure favour the cluster-cluster and 
cluster-group collisions, in the same way as the caustics seen in the 
simulations. 
\\
The most remarkable examples of cluster merging seen at an early  
stage are found in the central region of the Shapley Concentration, the 
richest supercluster of clusters within 300 \hmpc.
This supercluster represents a unique laboratory where it is possible to 
follow cluster mergings and to test related astrophysical consequences.
\\
We present the results of an extensive, multiwavelength survey of the 
central part of the Shapley Concentration, with the use of optical spectra,  
radio and X-ray data (see also the poster by Zucca et al., these proceedings).
\\
Updates to this project can be found at the WWW page:
\\
http://boas5.bo.astro.it/$\sim$bardelli/shapley/shapley\_new.html

%
\section{OPTICAL DATA}

We performed extensive redshift surveys in the central part of the Shapley
Concentration, covering both the cluster and the intercluster galaxies.
We focussed in particular on the A3558 and the A3528 complexes, two chains
of interacting clusters. We derived the dynamical parameters of each cluster
(Bardelli et al. 1994, 1998a, 2000b) and we studied their substructures
(Bardelli et al. 1998b, 2000b).
\\
The analysis of substructures is important in order to fully describe the
galaxy dynamics in major mergings: in this sense the DEDICA algorithm (Pisani 
1996) is particularly powerful, because it is possible to perform both a
bi-dimensional and a three-dimensional (i.e. using the galaxy velocity as
the third coordinate) analysis.
\\
In Figures 1 and 2 we show the results of this analysis on the A3528 and
the A3558 complexes, respectively.
\\
In the A3558 complex, the members of the chain appear to be fragmented in
a large number of components: in fact we found 21 significant three-dimensional 
subclumps. On the contrary, the A3528 complex is less fragmented:
in fact, even if these two complexes have similar sizes, in the A3528 complex
we detect only 8 significant three-dimensional subclumps.
 
%
\section{RADIO DATA}

We performed a radio survey at 22cm with the ATCA telescope to 1 mJy
in the region of the A3558 and A3528 cluster complexes, obtaining 
a total of 450 sources (Venturi et al. 2000a, 2000b). 
Among these sources, $\sim 100$ have optical   
counterpart and 43 have redshifts in the Shapley Concentration.
\\
In Figure 3 we show the detected radiosources overlaid on the optical 
isodensity contours: filled circles and squares are sources with
optical counterparts. The position of the radiosources at the distance
of the Shapley Concentration is reported in Figure 4, where the wedge diagrams
of the optical sample are presented. 
\\
The bivariate radio-optical lunimosity function has been calculated for
the A3528 and A3558 cluster complexes and compared with
the luminosity function of normal clusters (Ledlow \& Owen 1996) 
in Figure 5: it is clear that in the A3558 complex a 
significant lack of radiogalaxies is present. 
\\
In the region of interacting clusters, a number of extended
radiosources has been detected (Figure 6 and Figure 7). The study of 
these sources is particularly important for deriving physical constraints on 
the interaction of electrons with the ICM. In particular,
Venturi et al. (1999), studying a relic radio source at the periphery of
the A3558 complex, individuated a possible shock front between the
A3556 cluster and a smaller group.     
\\
Following this strategy we will integrate the optical, X-ray and radio data
in order to have a general description of the ongoing phenomena (Figure 8). 

%
\section{X-RAY DATA}

Fundamental data on the physics of the ICM during a merging come from the 
X-ray band. Figure 11 (taken from Ettori et al. 1997) shows a mosaic 
of ROSAT-PSPC images of clusters in the A3558 complex.
\\
Bardelli et al. (1996) and Kull \& B\"ohringer (1999) showed that the 
three clusters (A3562, A3558 and A3556) and the two poor groups (SC1329-131 and 
SC1327-312) of this complex are enbedded in a hot gas filament. 
This fact points toward a major merging event. 
A careful analysis of the distribution and the temperature of the ICM is 
currently in progress with the use of ROSAT and BeppoSAX data (Ettori et al. 
2000, see Figures 12 and 13). 
\\
For the A3528 cluster complex, pointed XMM observations have been allocated.  

%
\section{CONCLUSIONS}

In this contribution we described a multiwavelength study of two
cluster complexes at the center of the Shapley Concentration: both
these structures are formed by interacting clusters and have sizes of
the order of $\sim 7$ \hmpc. 

In the A3558 complex, the members of the chain appear to be fragmented in a 
large number of components, detected with high significance both in the optical 
and in the X-ray band. All these components are embedded in a hot gas filament. 
Moreover, Venturi et al. (2000) found a deficiency of radiogalaxies in the 
A3558 complex with respect to the radio-optical luminosity function of normal
clusters and Bardelli et al. (1998b) found an excess of blue galaxies in the
expected position of the shock.
\\
These facts could suggest that the dynamical processes acting on this complex
are in a rather advanced stage and that the merging events were already able to
induce modifications in the galaxy population.
Our conclusion is that the A3558 complex is a cluster-cluster collision
(otherwise called ``major merging") seen just after the first core-core
encounter, where an intervening cluster impacted onto the richer object
A3558. Indeed the clumpiness found eastward of A3558 could be due to the
galaxies of this intervening cluster, which is now emerging from the main
component. This scenario is also confirmed by numerical simulations of
merging clusters (see f.i. lower panels in Fig.2 of Burns et al. 1994,
which show the structure after the core-core encounter).

In the A3528 cluster complex, the effect of interactions is less spectacular.
This complex appear less fragmented than the A3558 complex and the bivariate 
radio-optical luminosity function is not different with respect to the other 
clusters.     
\\
All these characteristics of the A3528 complex seem to indicate that this
is a ``young" structure (Bardelli et al. 2000b): the two interacting cluster 
pairs (A3528N-A3528S and A3530-A3532) resemble the simulations of Burns et al. 
(1994) for the pre-merger case (see upper panels of their Fig.2). 
This scenario is consistent with the suggestions of Reid et al. (1998), based 
on the analysis of extended radiosources in this region and with the fact that 
the merging effects on the galaxy population and on the cluster dynamics are 
not yet evident.
\\
Given the overall overdensity of this region, at the end these two main
components will merge together in a major merging event, forming a structure
similar to the A3558 complex.

The ``age" difference of the two complexes is mainly due to their
overdensities, which lead to different collapse times (see Table 5 in
Bardelli et al. 2000a). This fact confirms that the Shapley Concentration
is a ``laboratory" where it is possible to study the formation of clusters
at different stages.

\begin{figure}
\begin{center}
\epsfysize=13cm
\epsfbox{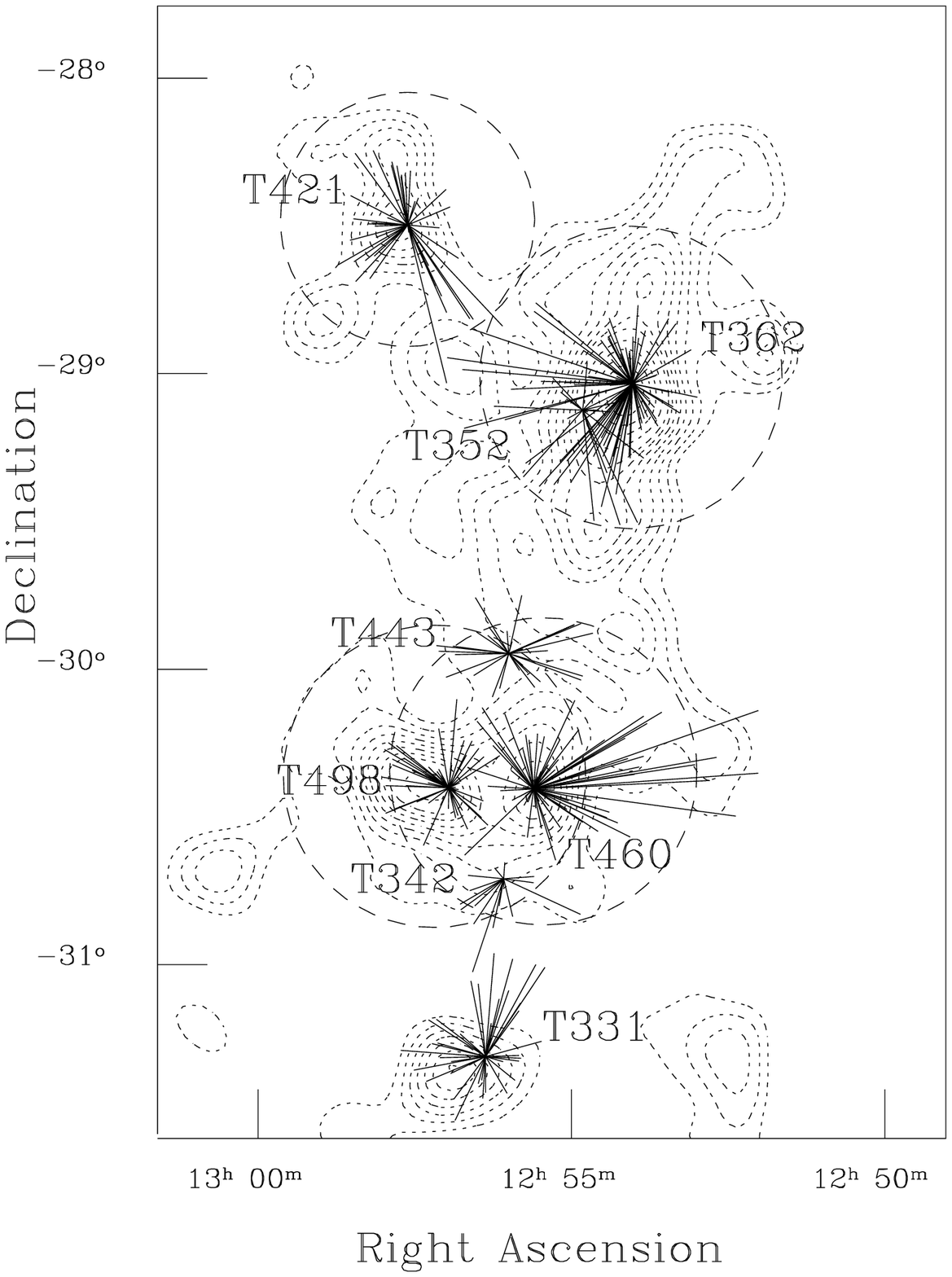}
\epsfysize=8.5cm
\epsfbox{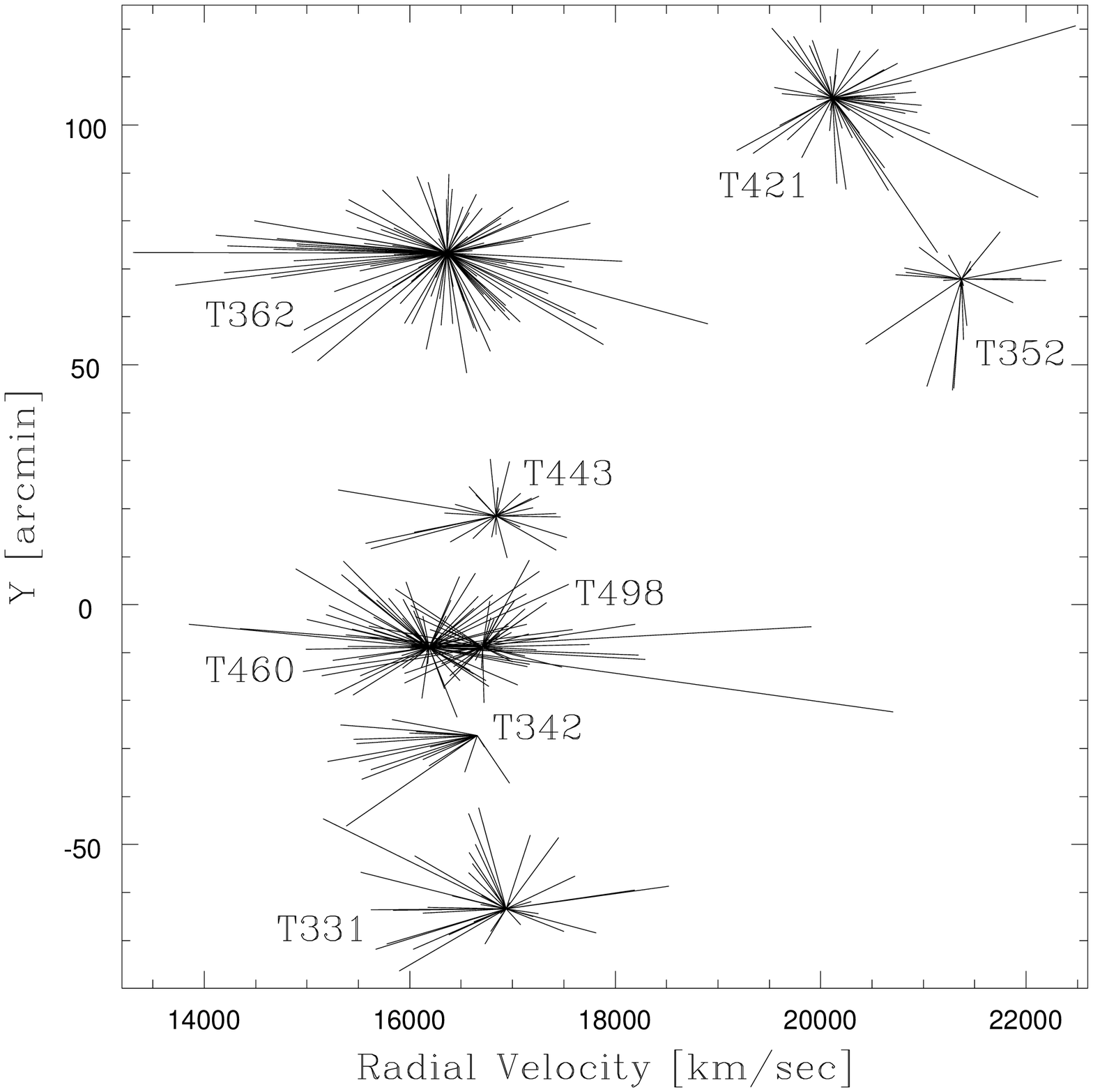}
\end{center}
\par\noindent
Figure 1: Substructures in the A3528 region found in the three--dimensional 
sample using the DEDICA algorithm.
Solid lines connect the position of each galaxy to the common limiting
position $\vec{x}_{lim}$ of the group. 
\\
Upper panel: projection on the plane of the sky. Lower panel: projection on
the velocity--Y plane.  
\end{figure}

\begin{figure}
\begin{center}
\leavevmode
\epsfxsize=0.49\hsize \epsfbox{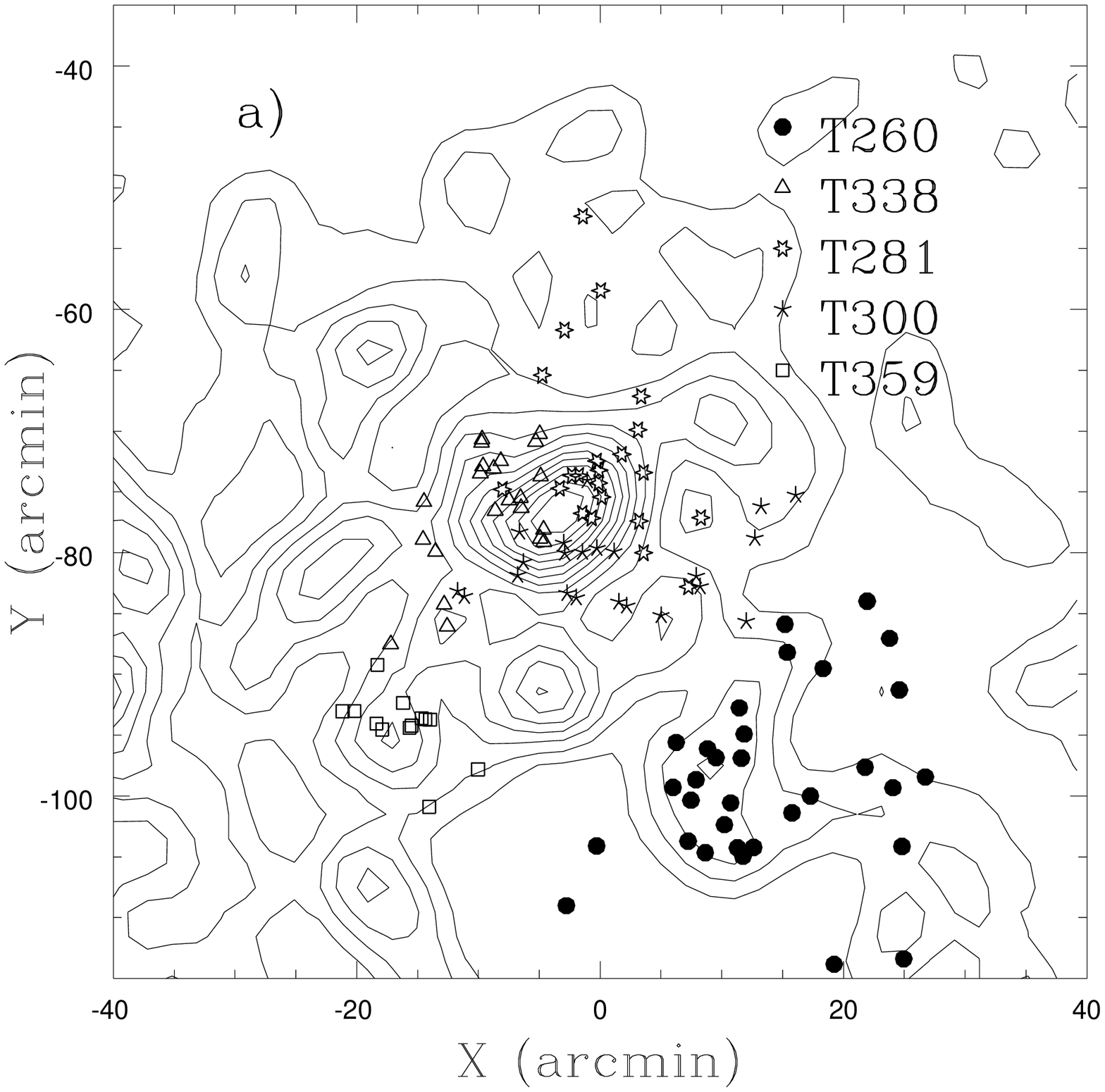} \hfil
\epsfxsize=0.49\hsize \epsfbox{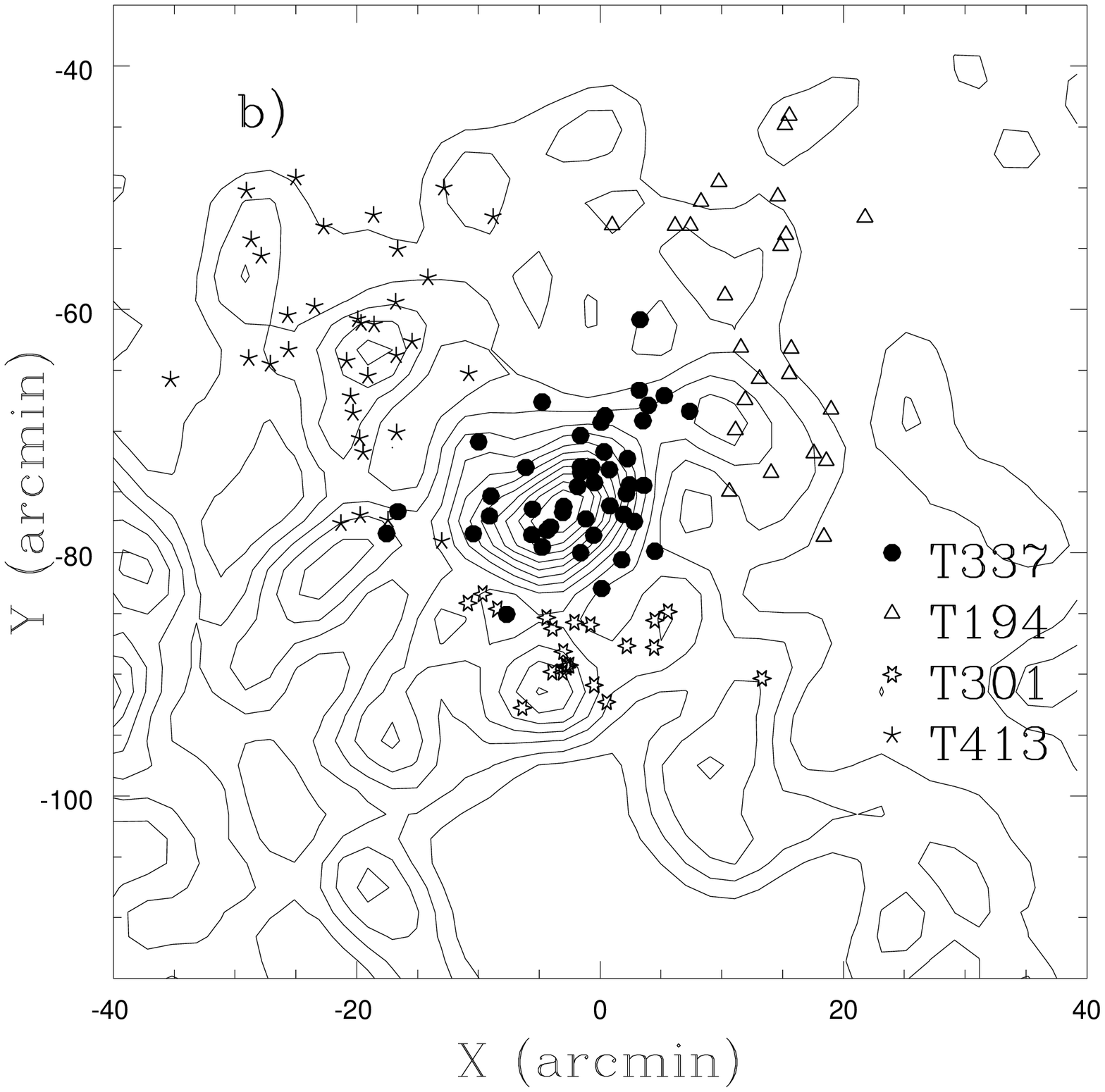} \hfil
\leavevmode
\epsfxsize=0.49\hsize \epsfbox{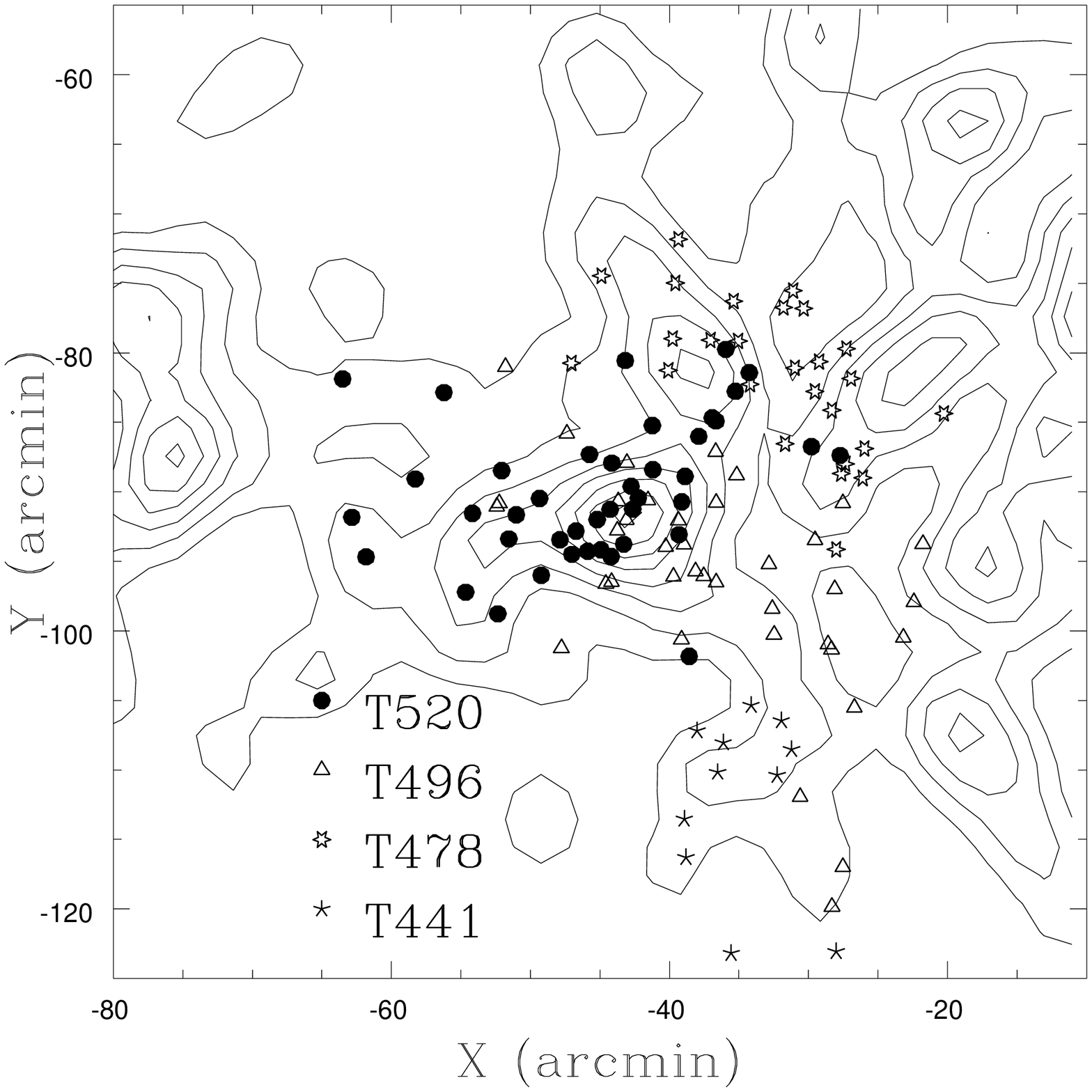} \hfil
\epsfxsize=0.49\hsize \epsfbox{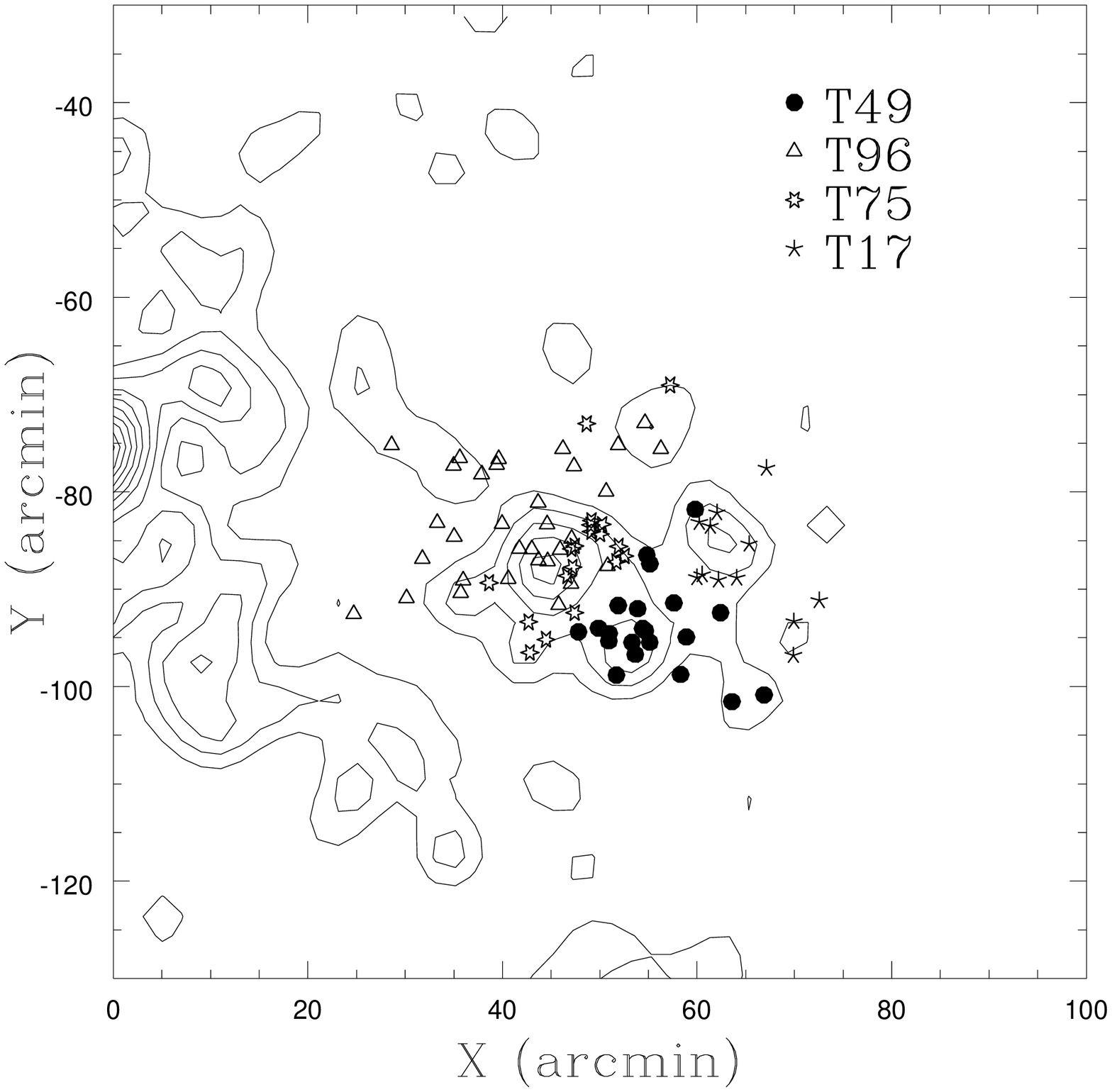} \hfil
\end{center}
\par\noindent
Figure 2: The positions of subcluster galaxy found by DEDICA in the 
three-dimensional sample covering the A3558 complex are overplotted on 
the 2D isodensity contours of galaxy distribution. 
The high number of substructures reveals that this structure is far from
the equilibrium state.  
\\
Upper panels: Substructures found in the A3558 cluster; for clarity, the
group members have been split in two panels.
\\
Lower panels: The same as above for SC1329-313 \& SC1327-312 (left) and A3556 
(right).
\end{figure}

\begin{figure}
\begin{center}
\epsfxsize= 0.8 \hsize
\epsfbox{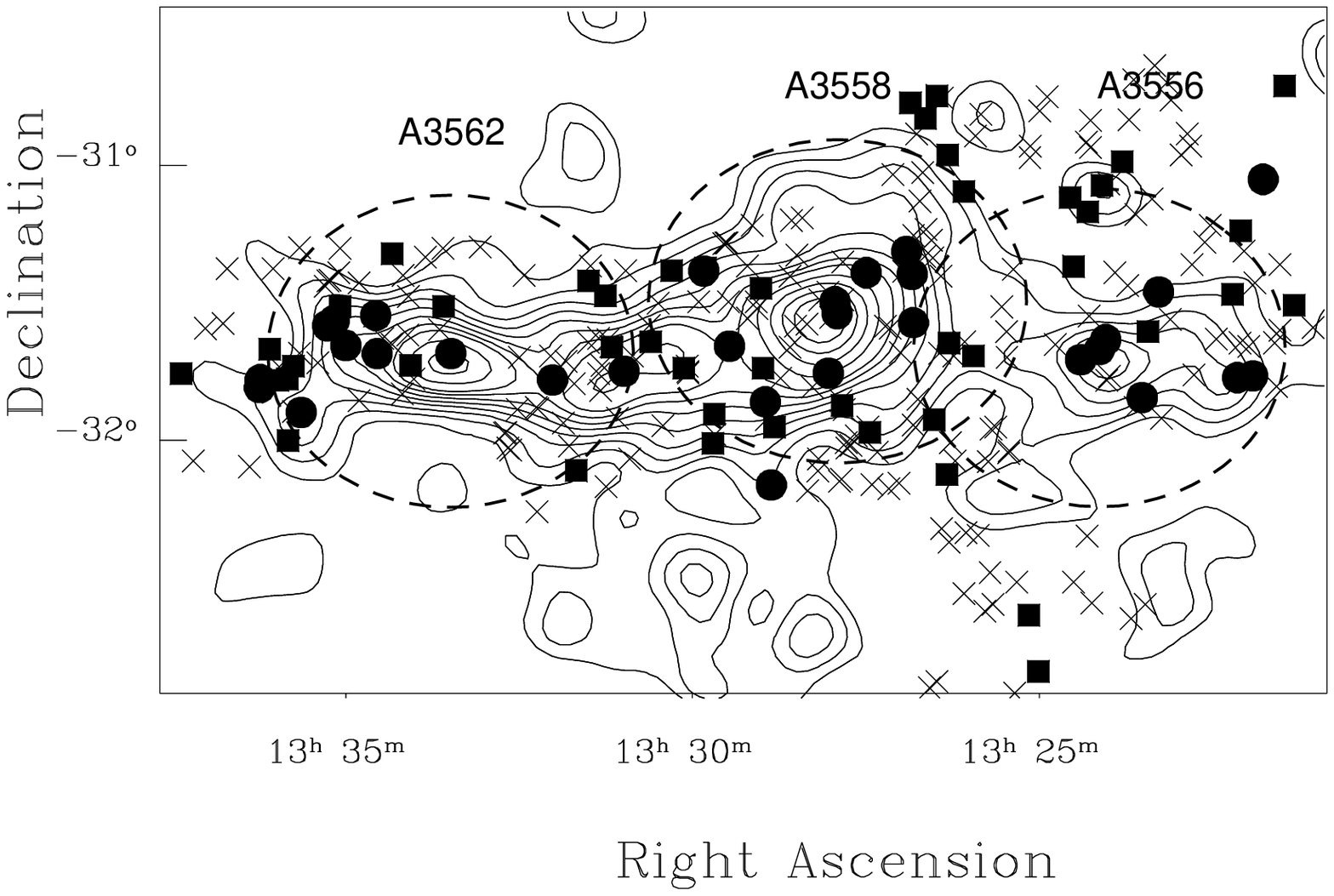}
\epsfxsize=0.75 \hsize  
\epsfbox{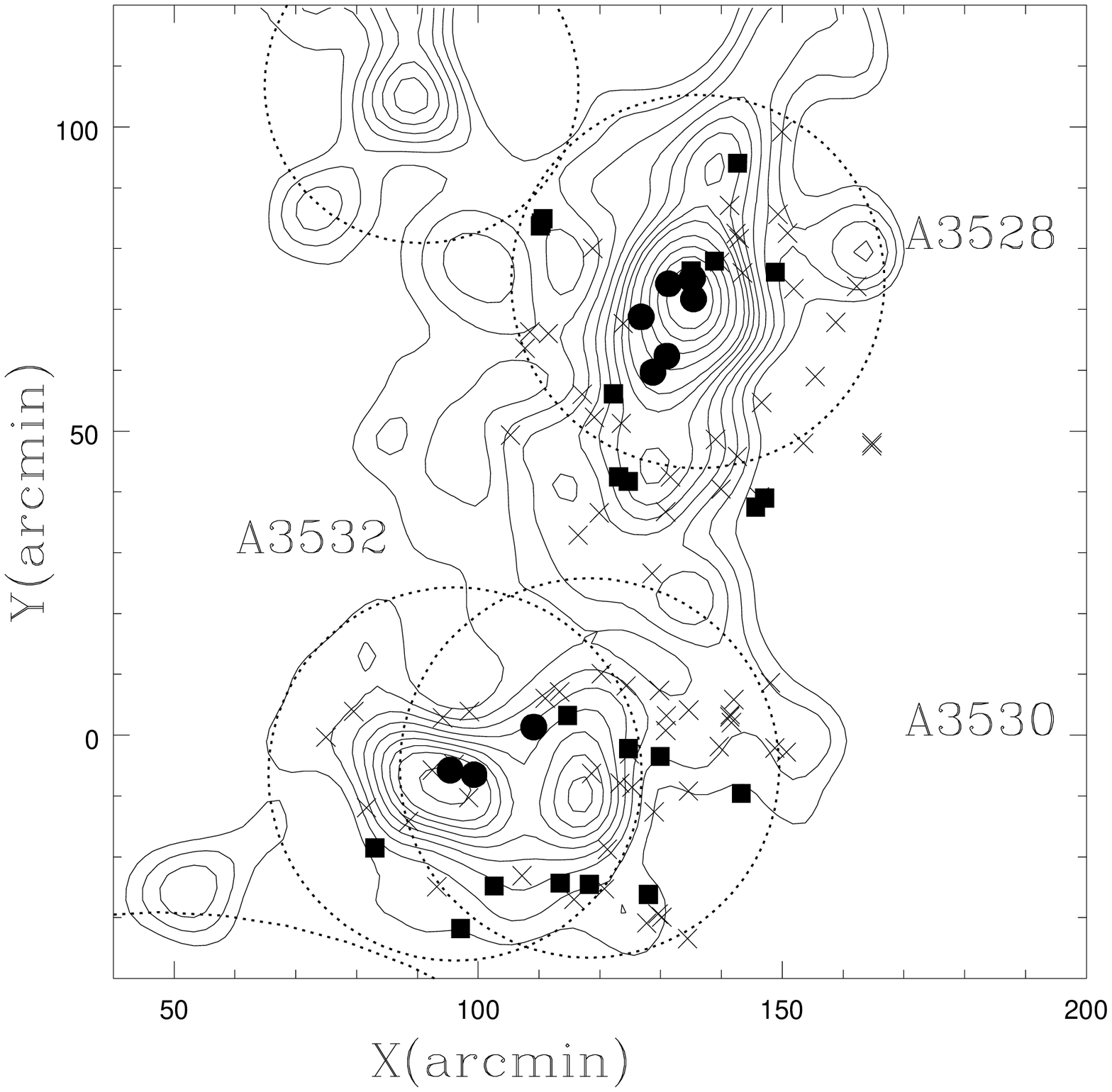}
\end{center}
\par\noindent
Figure 3: Radio sources detected in the 22cm ATCA survey in the A3558 and 
A3528 complex regions overlaid on the optical isodensities. Filled circles 
represent optical counterparts with measured redshift, filled squares represent
those without velocity information and crosses are radiosources without optical
counterparts. 
\end{figure}

\begin{figure}
\begin{center}
\epsfxsize= \hsize
\epsfbox{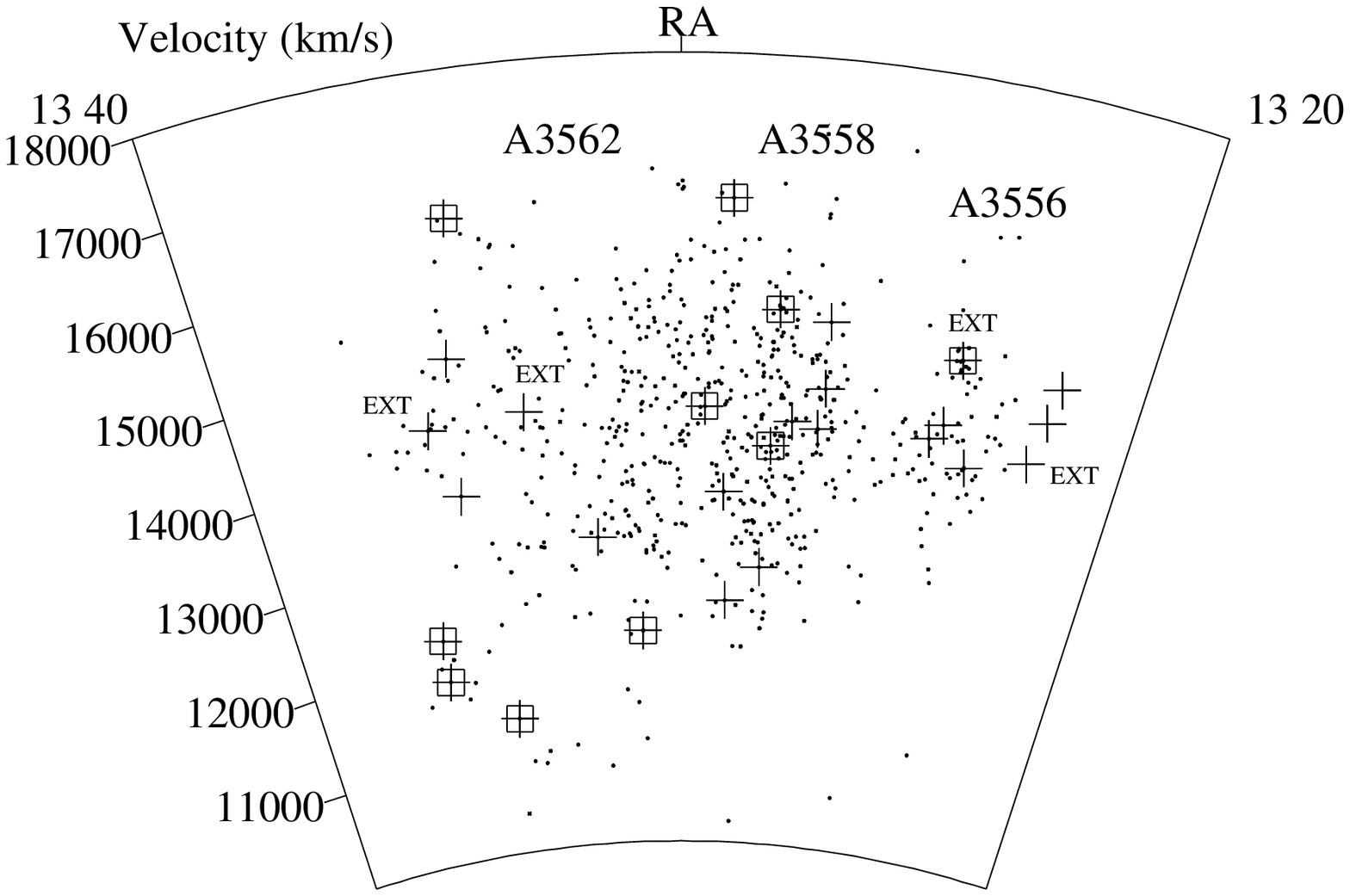}
\epsfxsize= 0.9 \hsize  
\epsfbox{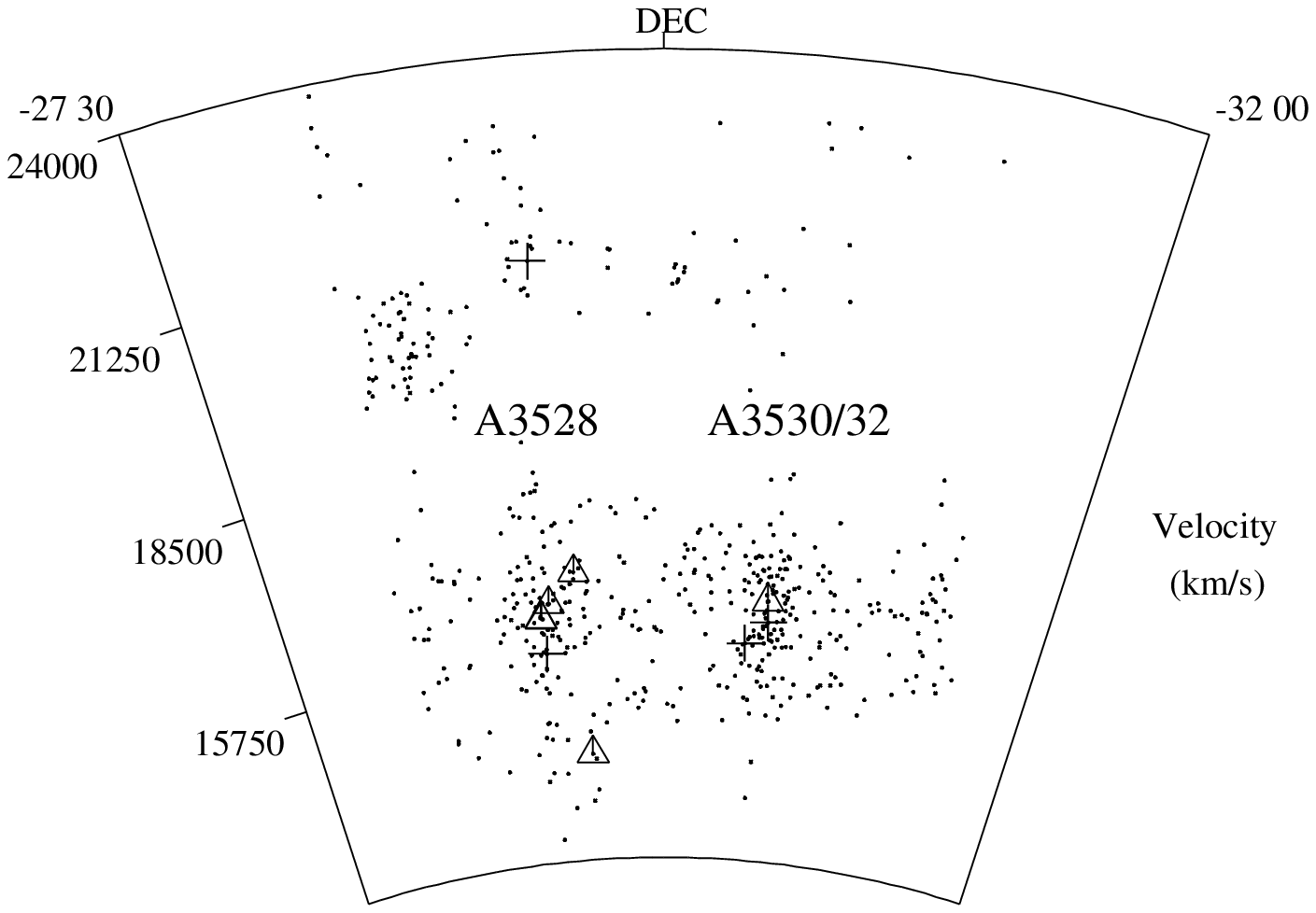}
\end{center}
\par\noindent
Figure 4: Wedge diagrams of galaxies in the A3558 and A3528 regions. Crosses
are radiosources with optical counterpart. In the A3558 complex, radio
galaxies with spectral index $\alpha^{22}_{13}>1$ have been marked with a 
square and the extended sources have been highlighted.
In the A3528 complex, extended radio galaxies are indicated by triangles. 
\end{figure}

\begin{figure}
\begin{center}
\epsfxsize=0.8 \hsize
\epsfbox{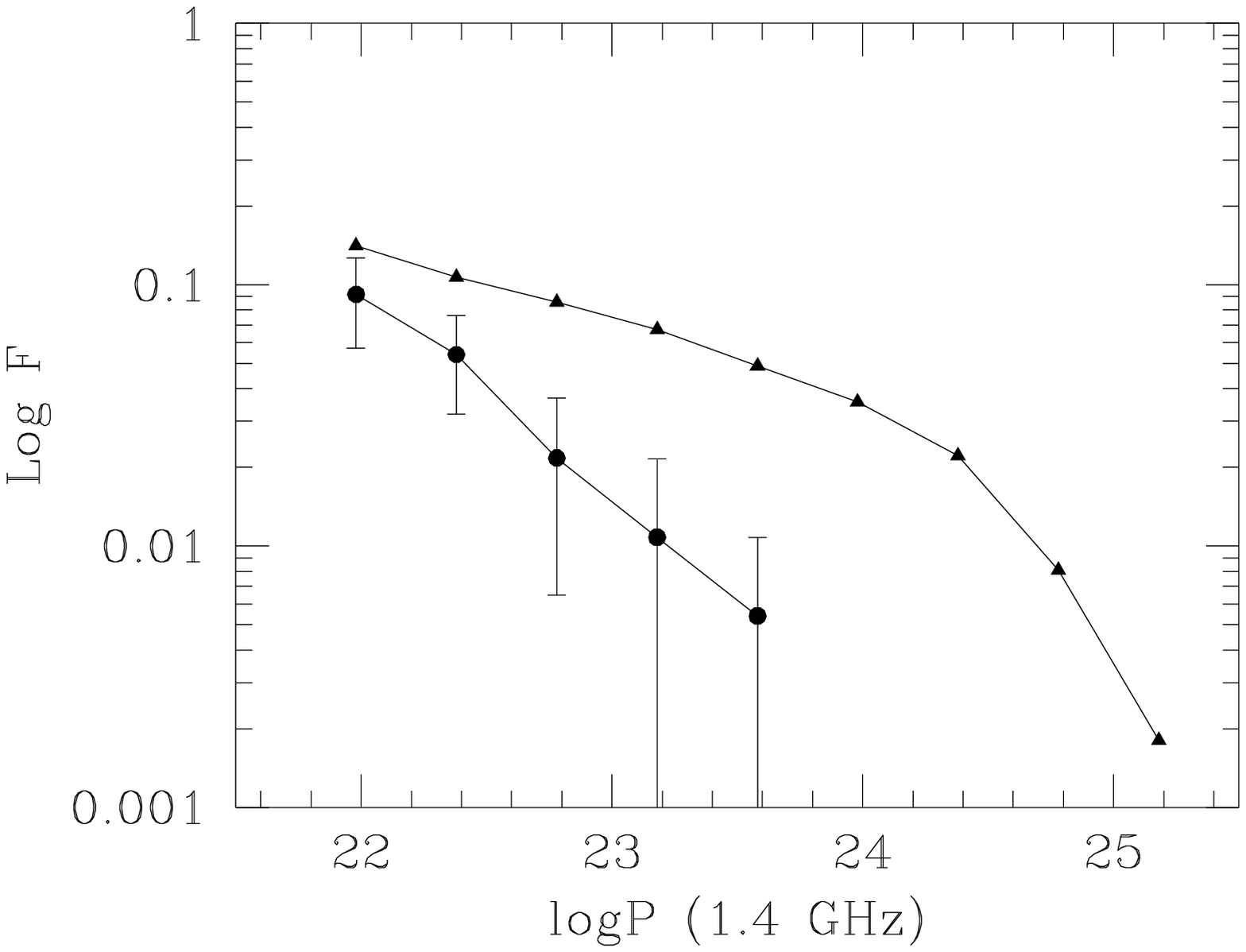}
\epsfxsize=0.65  \hsize  
\epsfbox{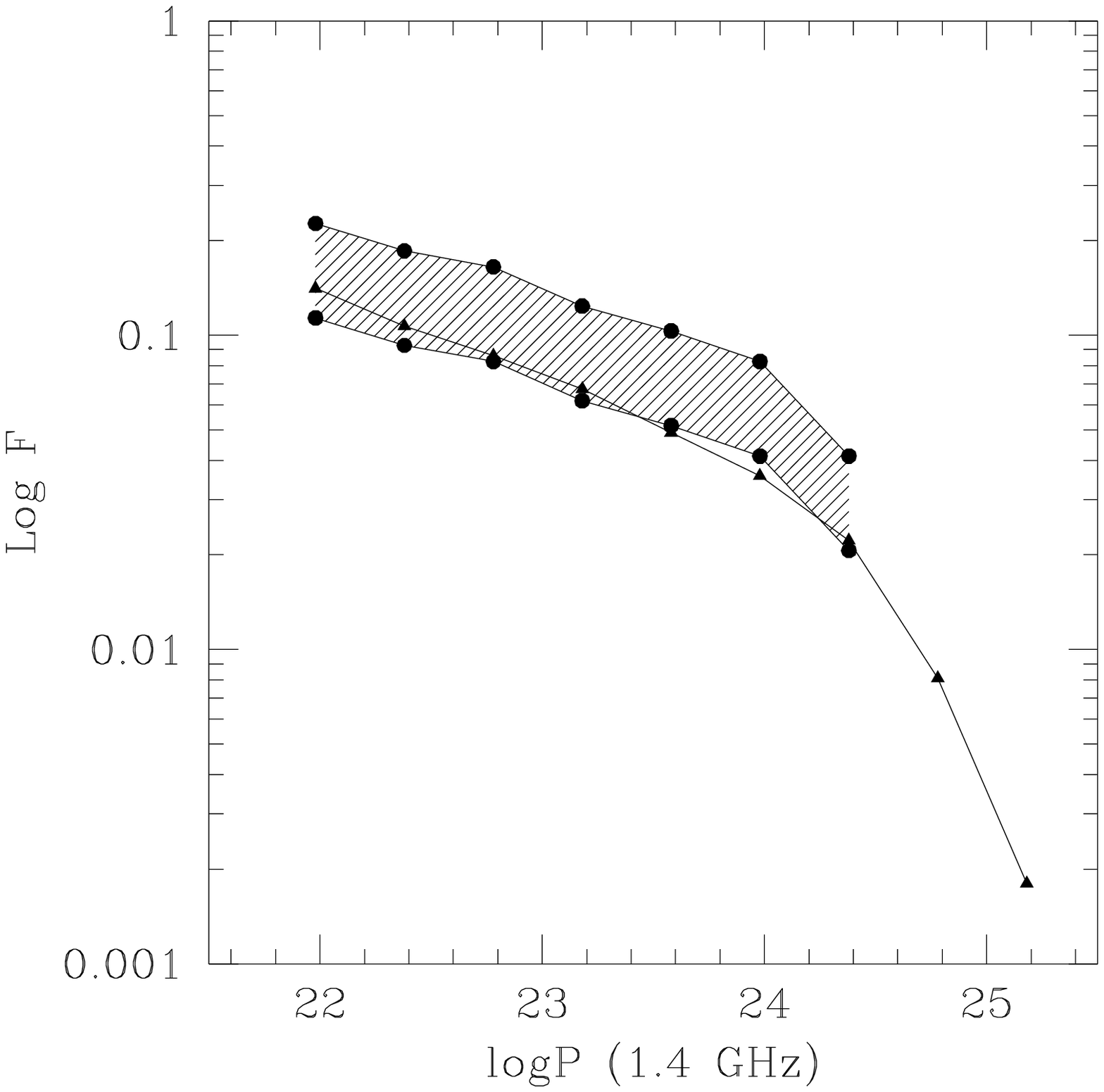}
\end{center}
\par\noindent
Figure 5: Bivariate radio-optical luminosity function for the A3558 (upper 
panel) and the A3528 (lower panel) complexes. Given the uncertainty in the 
fraction of ellipticals, the real luminosity function of the A3528 complex 
should lie in the shaded region.
Triangles represent the normal cluster luminosity function of 
Ledlow \& Owen (1996): note the lack of radio sources in the A3558 complex
with respect to normal clusters. 
\end{figure}

\begin{figure}
\begin{center}
\epsfxsize= \hsize
\epsfbox{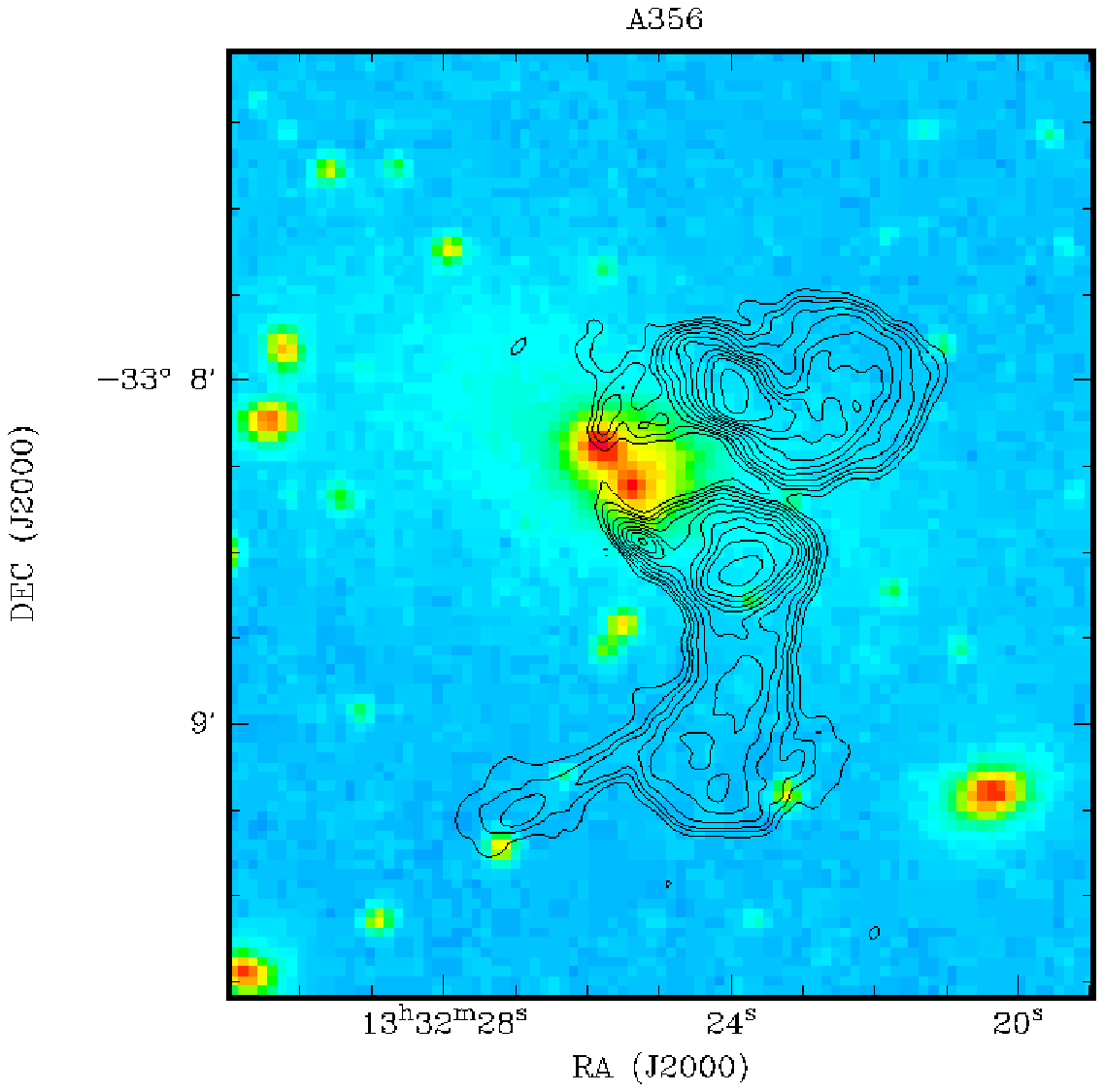} 
\end{center}
\bigskip
\par\noindent
Figure 6: Radio(20cm VLA)-optical superposition at the center of the cluster 
A3560. This kind of information, added to X-ray maps and spectra, will
constrain the dynamics of merging clusters. 
\end{figure}

\begin{figure}
\begin{center}
\epsfxsize=0.75 \hsize
\epsfbox{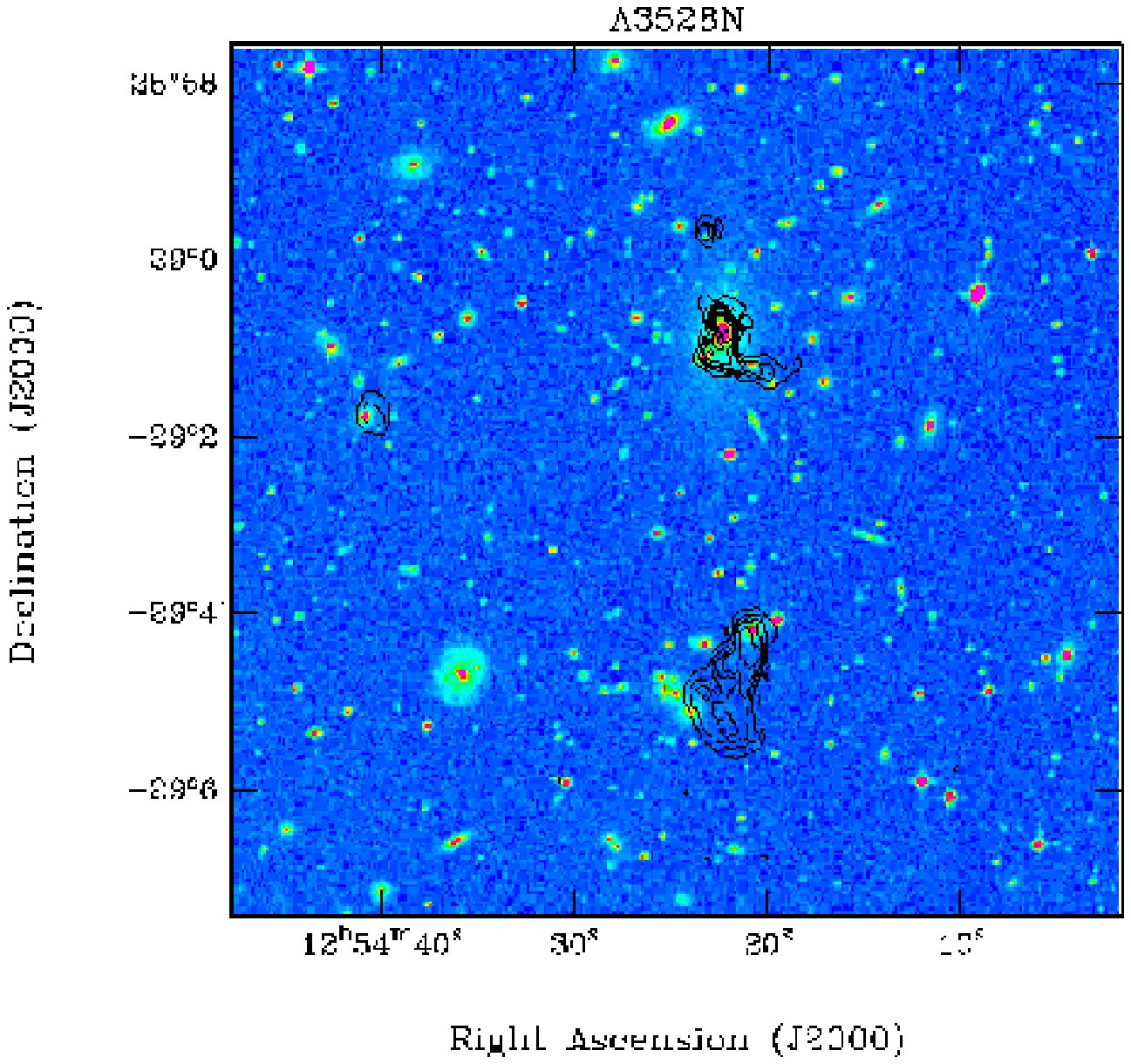}
\epsfxsize=0.75  \hsize  
\epsfbox{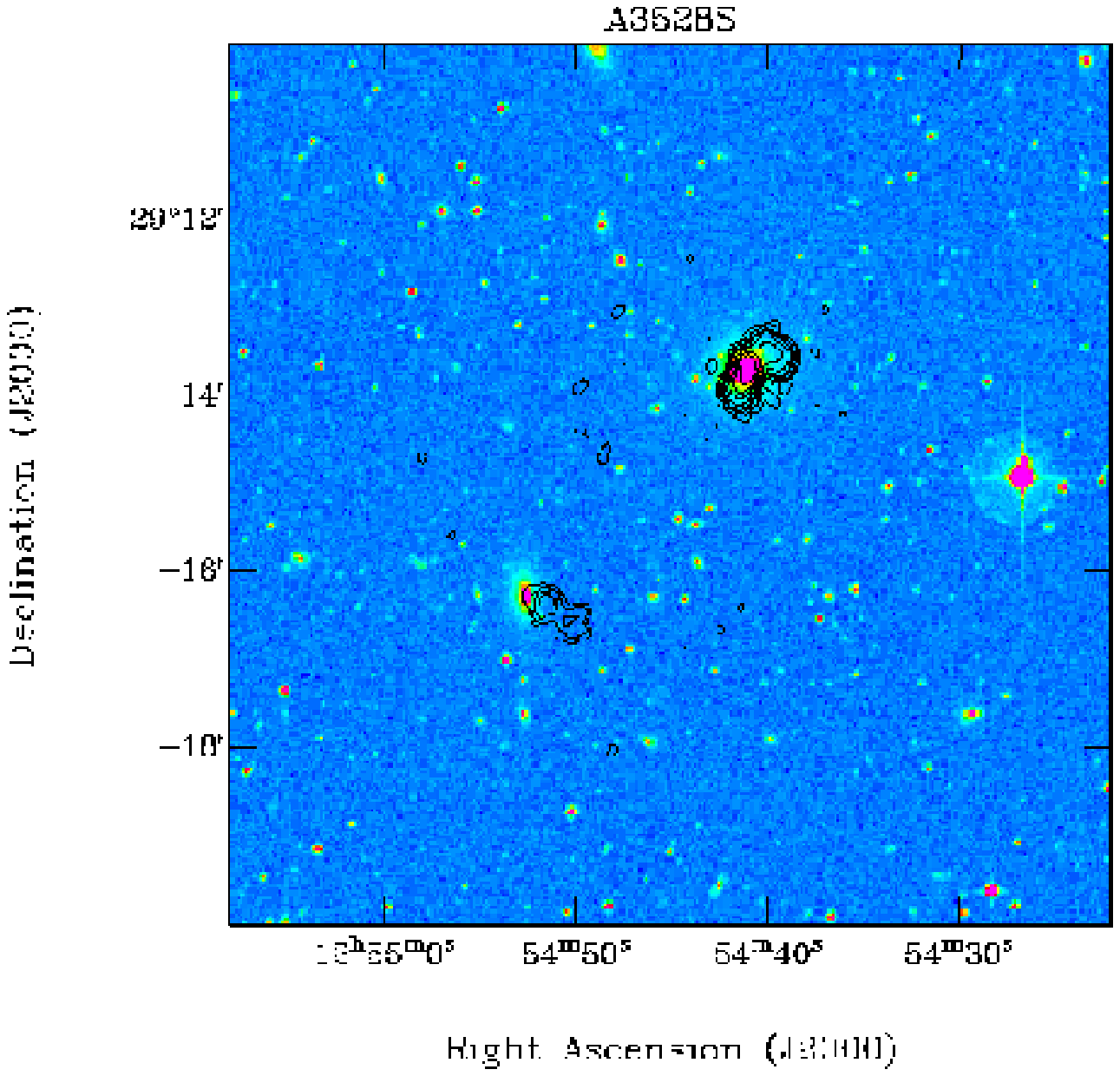}
\end{center}
\par\noindent
Figure 7: The A3528 cluster is actually a double cluster. In this figure
we superimposed the radio maps of the two components (dubbed A3528N and A3528S,
Schindler 1996) on the Digital Sky Survey. Note the central galaxy of A3528N, 
which is formed by two radiosources.   
\end{figure}

\begin{figure}
\begin{center}
\epsfxsize=0.9 \hsize
\epsfbox{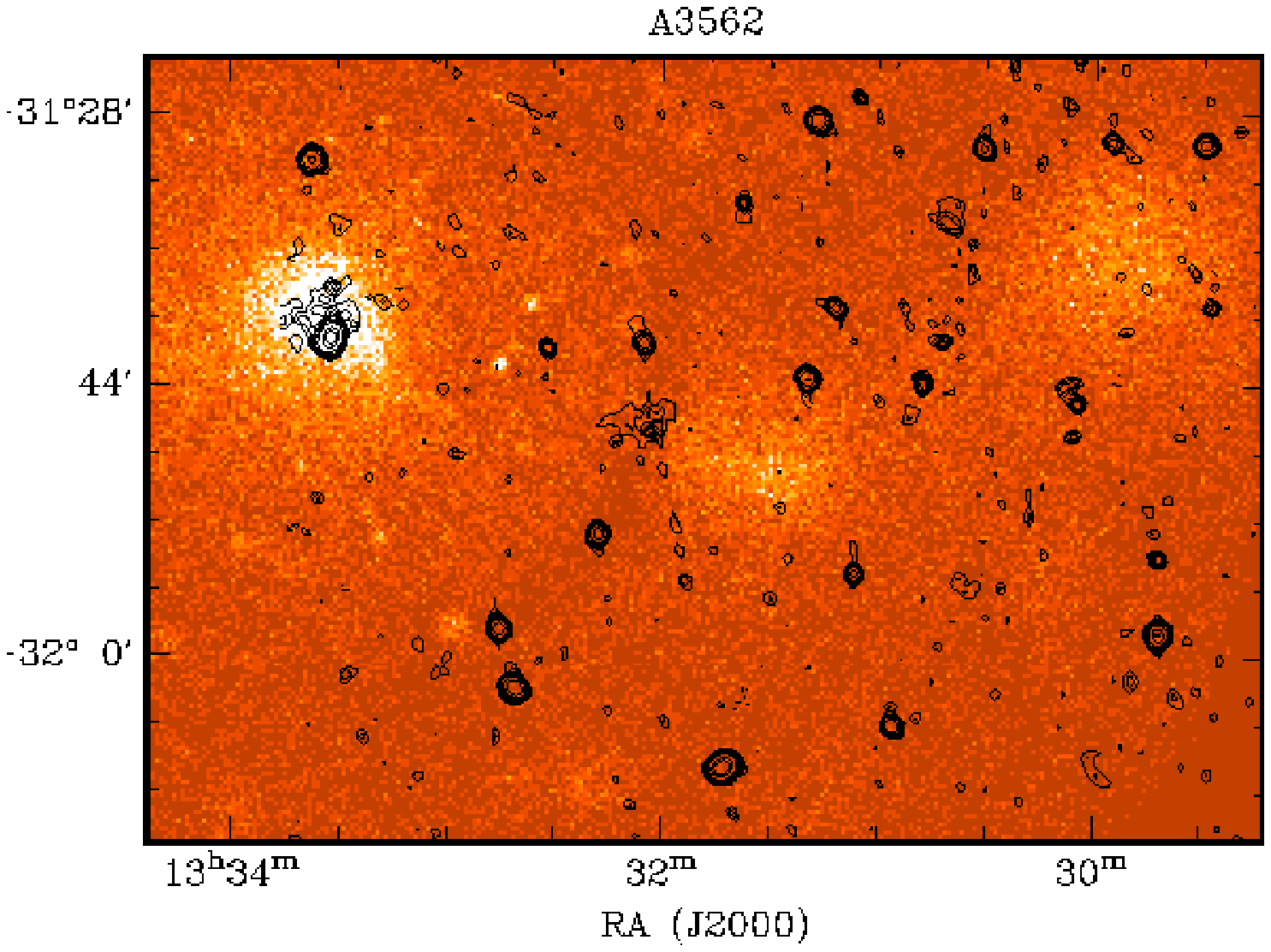}
\end{center}
\bigskip 
\par\noindent
Figure 8: The radio isophotes from the NVSS survey are superimposed to a part 
of the ROSAT PSPC image centered on A3562. 
At the center of the cluster A3562 (on the left) the emission of an Head Tail 
radio galaxy is clearly visible.
Additional extended emission is present in the cluster center. At the center
of the figure, an extended radio emission (possibly a mini-halo) is located at 
the border of the poor cluster SC 1329-313. This emission is associated to a
bright cluster galaxy. 
\end{figure}

\begin{figure}
\begin{center}
\epsfxsize= \hsize 
\epsfbox{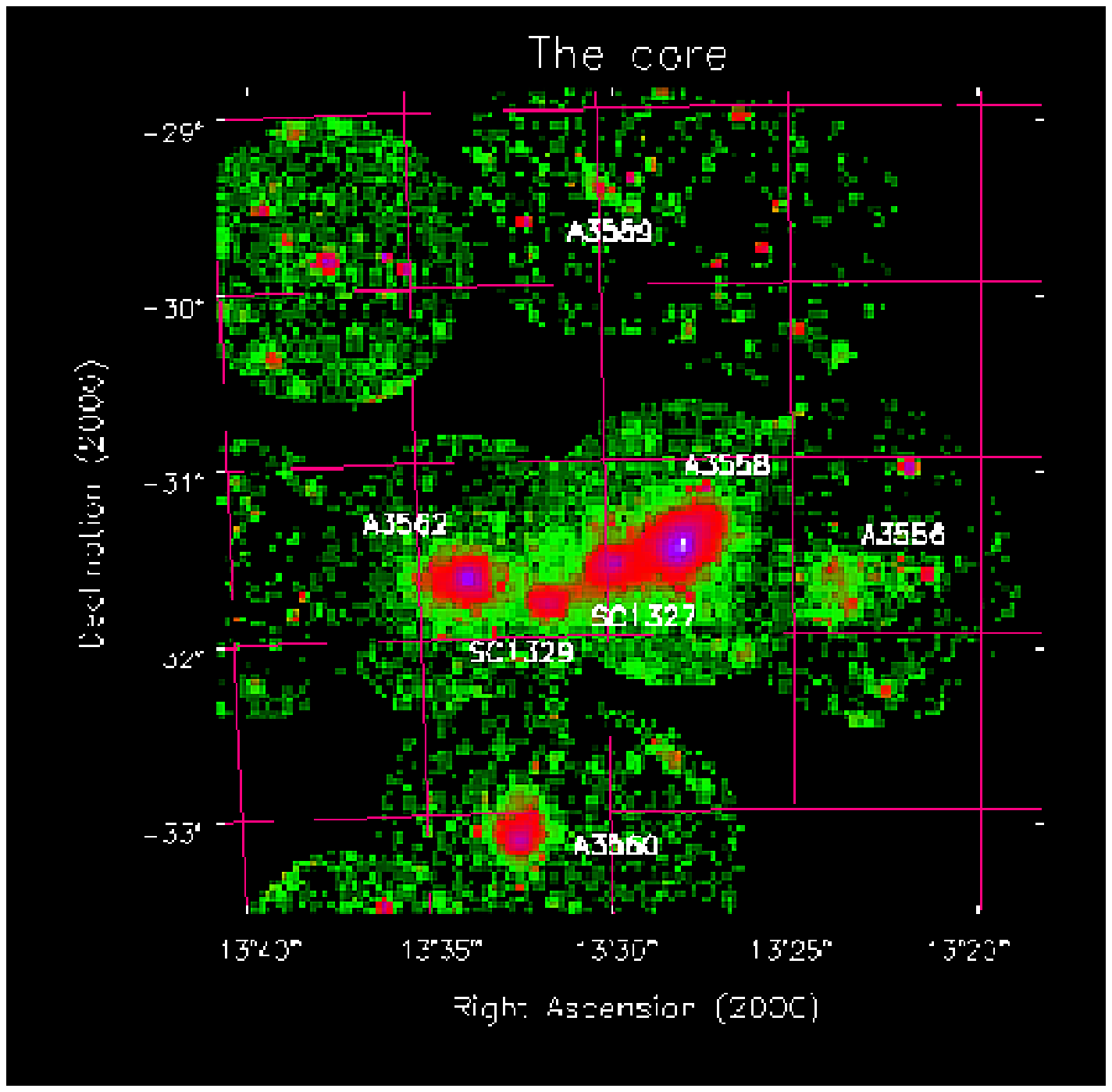}
\end{center}
\par\noindent
Figure 9: The soft X-ray emission (ROSAT mosaic from Ettori et al. 1997)
in the A3558 cluster complex, with the main clusters labelled.
We propose that this structure is the remnant of a cluster-cluster collision
seen after the first core-core encounter.
\end{figure}

\begin{figure}
\begin{center}
\epsfxsize=0.7 \hsize
\epsfbox{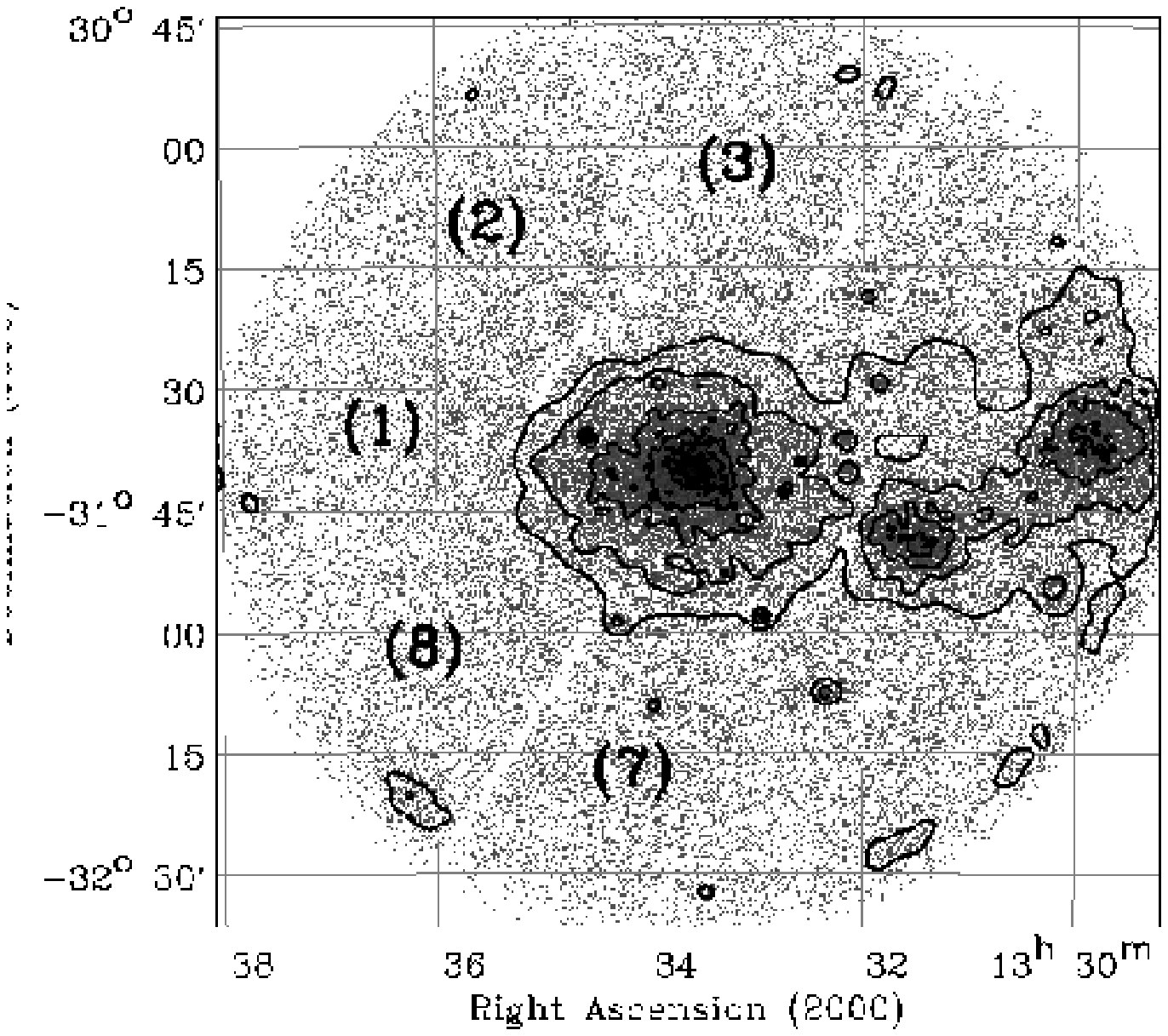}
\epsfxsize=0.7 \hsize
\epsfbox{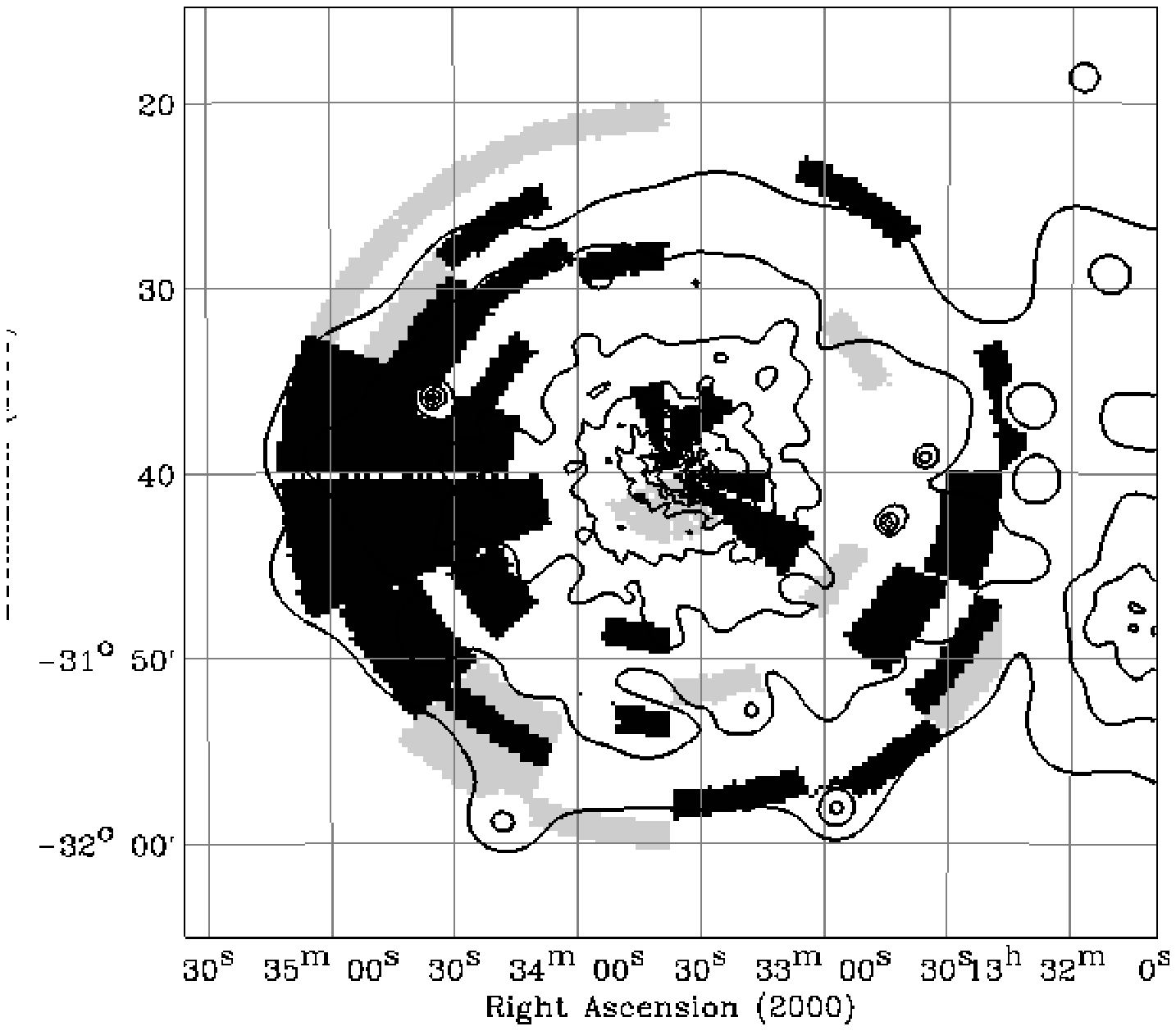}
\end{center}
\par\noindent
Figure 10: Upper panel: ROSAT PSPC image of the A3562 cluster. The image has 
been adaptively smoothed in order to enhance the features with a significance 
higher than 5 $\sigma$ with respectc to the background. Lower panel:
Residuals in the emission map after subtraction of a best fit $\beta$ model.
The darker regions represent excess in emission, the lighter ones indicate a 
deficit.
\end{figure}

\begin{figure}
\epsfxsize=0.7 \hsize
\begin{center}
\rotate[r]{\epsfbox{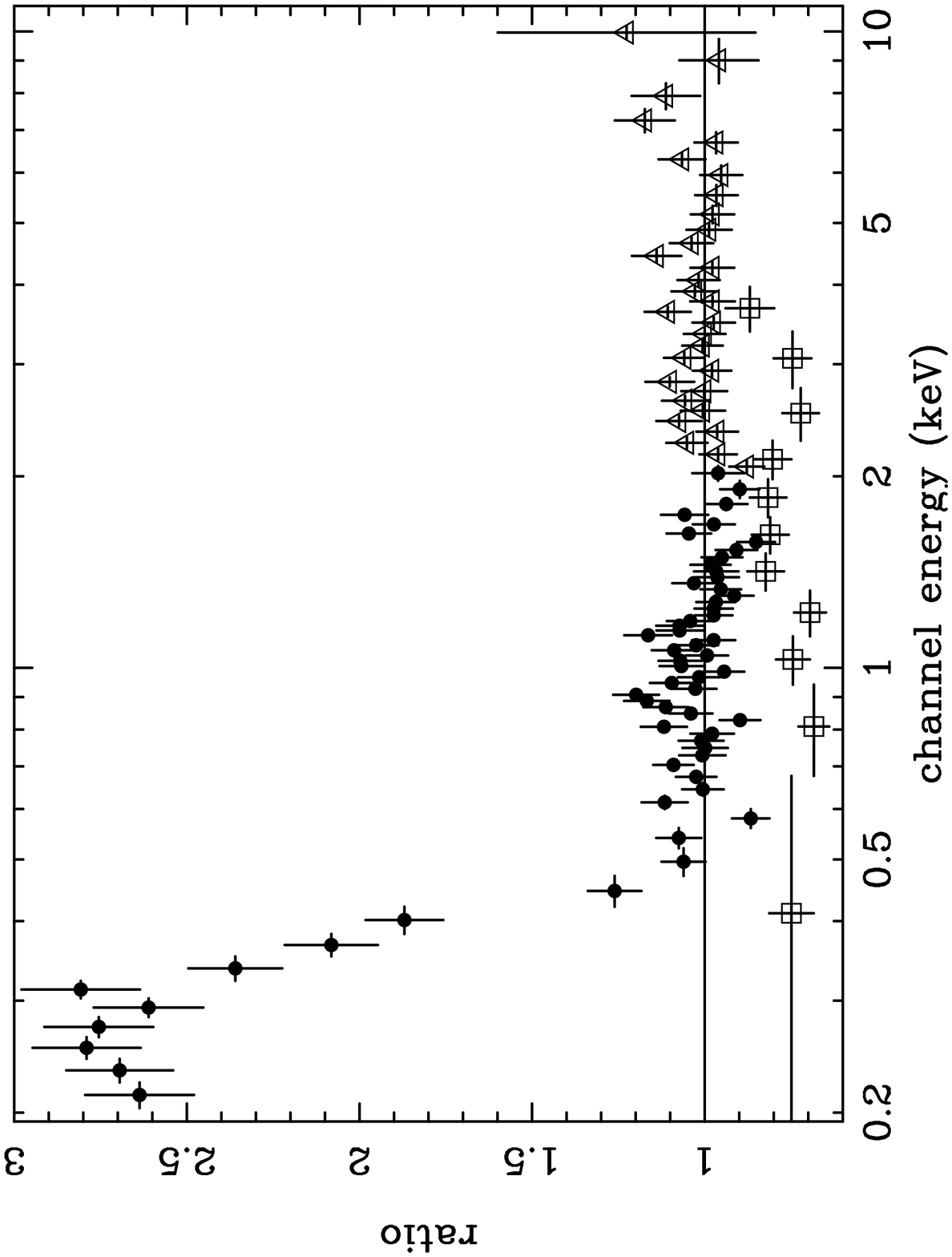}}
\epsfxsize=0.7 \hsize
\epsfbox{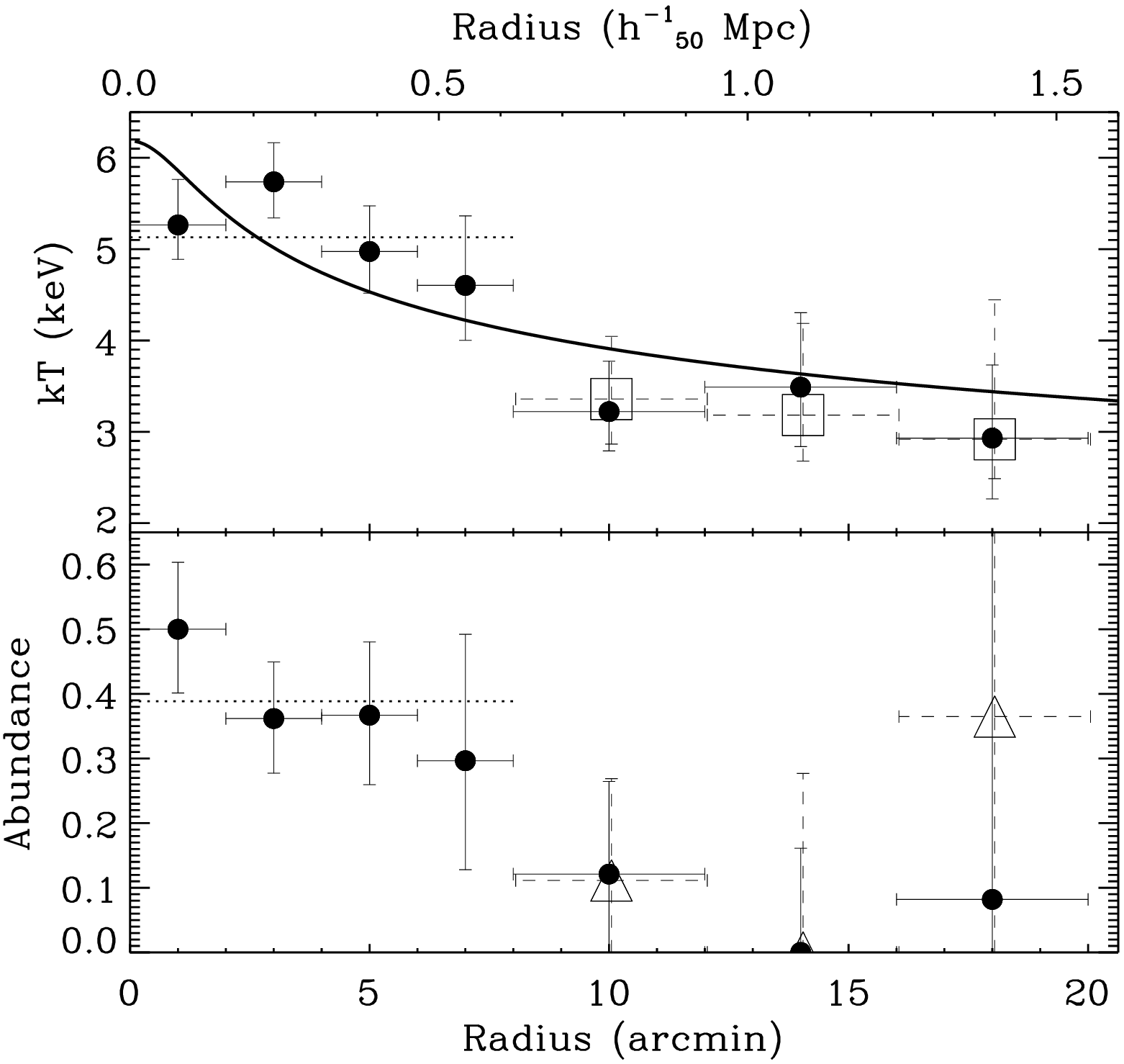}
\end{center}
\par\noindent
Figure 11: Upper panel: The LECS (open squares) and PSPC (filled circles) 
counts of A3562 are here divided by the best-fit absorbed MEKAL model with
fixed galactic absorption obtained from the MECS data (open triangles).  
Note that the LECS data are shifted by a factor $\sim 0.77$ in normalization 
with respect to MECS and that the PSPC counts present a large disagreement
below 0.5 keV.
Lower panel: Projected radial temperature and abundance profiles for A3562,
as determined from the BeppoSAX map. Filled circles and open triangles indicate
the measurements obtained including and excluding point sources. 
\end{figure}

\begin{figure}
\begin{center}
\epsfxsize= 0.7 \hsize
\epsfbox{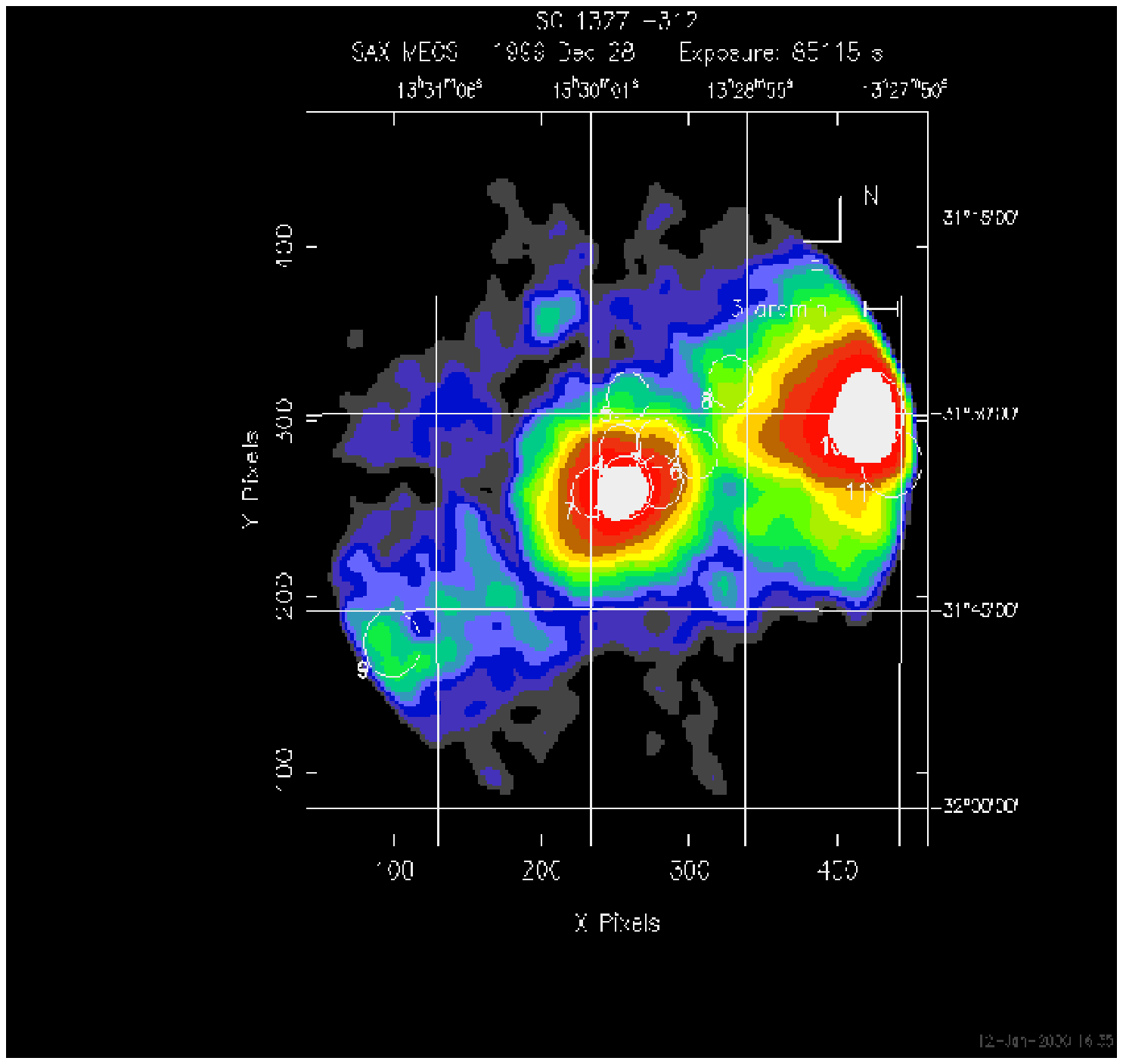}
\epsfxsize=0.7 \hsize  
\epsfbox{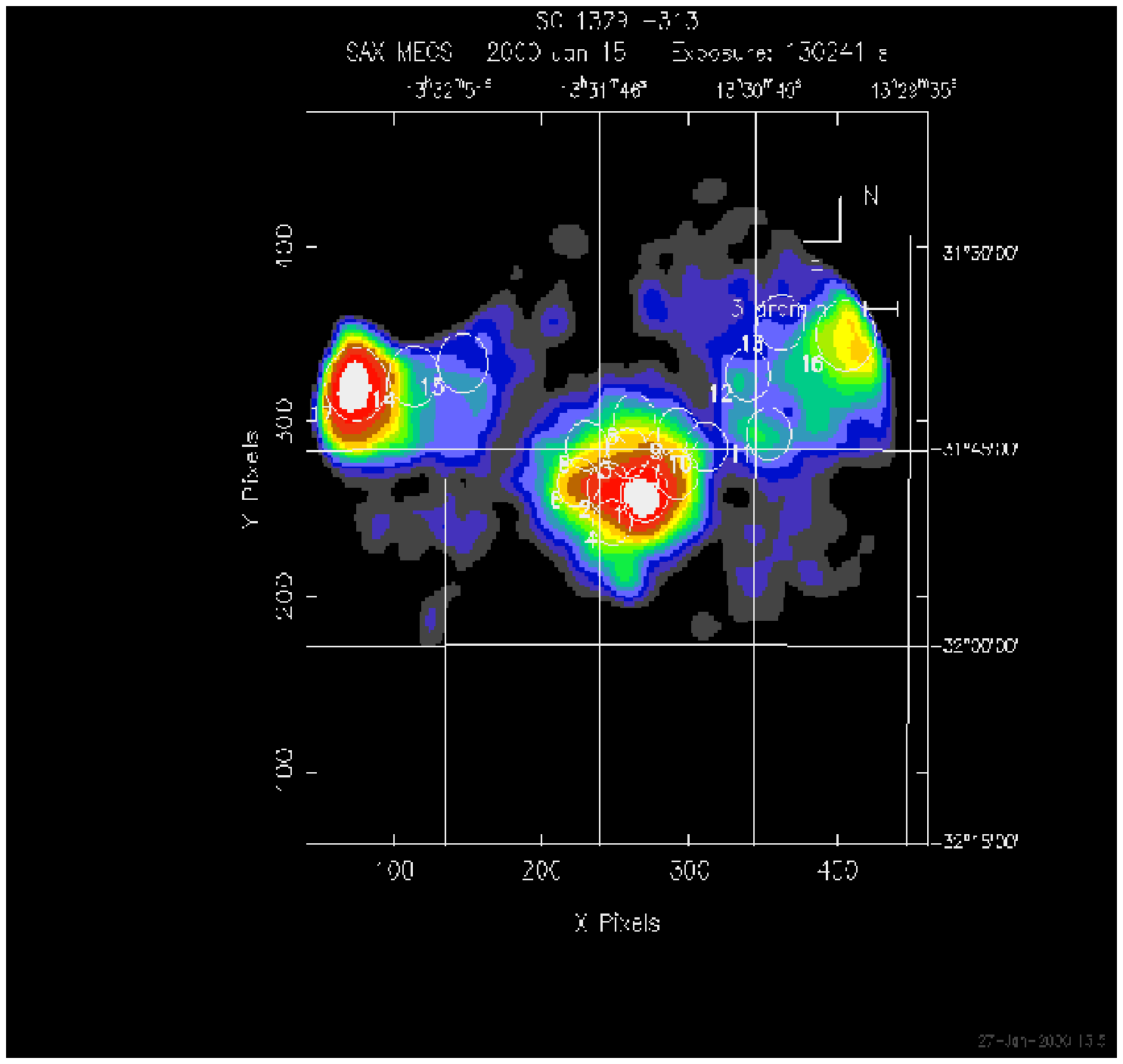}
\end{center}
\par\noindent
Figure 12: BeppoSAX MECS maps of the two poor groups SC1327-312 and SC1329-313, 
which are in between A3558 and A3562. 
\end{figure}

%


\section*{References}

\begin{thebibliography}{99}

\bibitem{} Baldi A., Bardelli S., Zucca E., 2000, MNRAS submitted

\bibitem{} Bardelli S., Zucca E., Vettolani G., Zamorani G., Scaramella R., 
           Collins C.A., MacGillivray H.T., 1994, MNRAS 267, 665 

\bibitem{} Bardelli S., Zucca E., Malizia A., Zamorani G., Scaramella R.,
           Vettolani G., 1996, A\&A 305, 435 

\bibitem{} Bardelli S., Zucca E., Zamorani G., Vettolani G., Scaramella R., 
           1998a, MNRAS 296, 599

\bibitem{} Bardelli S., Pisani A., Ramella M., Zucca E., Zamorani G., 
           1998b, MNRAS 300, 589 

\bibitem{} Bardelli S., Zucca E., Zamorani G., Moscardini L., Scaramella R., 
           2000a, MNRAS 312, 540

\bibitem{} Bardelli S., Zucca E., Baldi A., 2000b, MNRAS in press
           (astro-ph/0009008)

\bibitem{} Burns J.O., Roettiger K., Ledlow M., Klypin A., 1994, 
           ApJL 427, L87
 
\bibitem{} Ettori S., Fabian A.C., White D.A., 1997, MNRAS 289, 787  

\bibitem{} Ettori S., Bardelli S., DeGrandi S., Molendi S., Zamorani G., 
           Zucca E., 2000, MNRAS in press

\bibitem{} Kull A., B\"ohringer H., 1999, A\&A 341, 23

\bibitem{} Ledlow M.J., Owen F.N., 1996, AJ 112, 9 

\bibitem{} Pisani A., 1996, MNRAS 278, 697 

\bibitem{} Reid A.D., Hunstead R.W., Pierre M.M., 1998, MNRAS 296, 949 

\bibitem{} Schindler S., 1996, MNRAS 280, 309 

\bibitem{} Venturi T., Bardelli S., Morganti R., Hunstead R.W., 1997, 
           MNRAS 285, 898 

\bibitem{} Venturi T., Bardelli S., Morganti R., Hunstead R.W., 1998, 
           MNRAS 298, 1113

\bibitem{} Venturi T., Bardelli S., Zambelli G., Morganti R., Hunstead R.W., 
           1999, in {\it Diffuse thermal and relativistic plasma in
           galaxy clusters}, H.B\"ohringer et al. eds., MPE report 271 p.27

\bibitem{} Venturi T., Bardelli S., Morganti R., Hunstead R.W., 2000a,
           MNRAS 314, 594 

\bibitem{} Venturi T., Bardelli S., Zambelli G., Morganti R., 
           Hunstead R.W., 2000b, MNRAS submitted 

\end{thebibliography}
\end{document}